# DYNAMIC AND ISOTOPIC EVOLUTION OF ICE RESERVOIRS ON MARS

E. Vos[1], O. Aharonson[1,2], N. Schorghofer[2]

[1]Department of Earth and Planetary Sciences, Weizmann Institute of Science, Rehovot, Israel 76100

[2]Planetary Science Institute, Tucson, AZ 85719, USA

*Contact Information: e-mail: Eran.Vos@weizmann.ac.il

**Key Points**

1. We link Mars's orbital elements with the stratigraphy and isotopic composition of its ice by modeling the exchange among its reservoirs.

2. The precession period of 50 kyr dominates the isotopic composition during epochs of low and nearly constant obliquity such as at present.

3. Isotopic sampling of the top 100 meters may reveal climate oscillations unseen in the layer thicknesses.

**Abstract**

The layered polar caps of Mars have long been thought to be related to variations in orbit and axial tilt. We dynamically link Mars's past climate variations with the stratigraphy and isotopic composition of its ice by modeling the exchange of $H_2O$ and HDO among three reservoirs. The model shows that the interplay among equatorial, mid-latitude, and north-polar layered deposits (NPLD) induces significant isotopic changes in the cap. The diffusive properties of the sublimation lags and dust content in our model result in a cap size consistent with current Mars. The layer thicknesses are mostly controlled by obliquity variations, but the precession period of 50 kyr dominates the variations in the isotopic composition during epochs of relatively low and nearly constant obliquity such as at present. Isotopic sampling of the top 100 meters may reveal climate oscillations unseen in the layer thicknesses and would thus probe recent precession-driven climate cycles.

**Introduction**

Beyond the conspicuous polar layered deposits (Byrne, 2009; Laskar et al., 2002; Phillips et al., 2008), ice is found on Mars buried by a thin cover in the mid-latitudes (Byrne et al., 2009; Feldman et al., 2002) and more deeply in ancient equatorial glaciers (Head and Marchant, 2003; Head et al., 2005; Levrard et al., 2007; Levy et al., 2014; Shean et al., 2005). The PLD on Mars are widely believed to harbor a record of past climatic oscillations driven by variations in orbital elements (Byrne, 2009; Cutts and Lewis, 1982; Laskar et al., 2002). However, establishing a mechanistic link between orbital variations, climate, and the observed stratigraphy has proven elusive. Previous work suggested that accumulation rates are correlated with orbital parameters (Cutts and Lewis, 1982; Hvidberg et al., 2012; Laskar et al., 2002; Levrard et al., 2007; Milkovich and Head, 2005; Phillips et al., 2008). On Earth, isotopic ratios in ice cores reveal past temperature (Alley, 2014), but Earth's water is sourced from an ocean of constant D/H ratio, so analogues models do not apply to Mars.

Here we present a three-box model (Figure 1), accounting for the primary reservoirs: the North Polar Layered Deposits (NPLD), mid-latitude subsurface deposits (SSD), and near-equatorial glacial deposits (GD). In this simple model, we neglect the role of other reservoirs, such as vapor sourced from the Southern hemisphere (at least to the extent that such sources may not be regarded as well-mixed with the GD). Tracking the fluxes exchanged among these reservoirs enables quantitative estimates of their development, and predicts the isotopic record in the stratigraphy. In our model, fractionation occurs due to temperature differences between accumulating reservoirs (Merlivat and Nief, 1967) (see Methods). Thus with these assumption, the simplest model that can exhibit fractionation requires at least three reservoirs.

Although both poles are capped by ice deposits, here we focus on the NPLD as the larger reservoir, composed of exchangeable water ice and a few percent dust. Dust and dust-rich lag layers in the NPLD act to protect the ice from sublimation, both as diffusion barriers and thermal insulators (Hudson et al., 2007; Levrard et al., 2007). A second reservoir corresponds to mid-latitudes ice that has been observed to reside in the subsurface by gamma-ray and neutron spectroscopy (Feldman et al., 2002; Mitrofanov et al., 2002), the Phoenix Lander (Mellon et al., 2009; Smith et al., 2009) and in fresh craters exposing the subsurface (Byrne et al., 2009). This ice is stable in mid to high latitudes where a dry layer acts as thermal insulator (Schorghofer and Aharonson, 2005) to protect it from the diurnal and seasonal temperature variations (Levrard et al., 2004; Mellon et al., 2009; Mellon et al., 2004). Ice loss from the SSD is dominantly from a narrow latitude range in the mid-latitudes. The contribution from changes in ice table depth at higher latitudes is volumetrically smaller (Schorghofer and Forget, 2012).

Lastly, geologic evidence points to a significant near-equatorial reservoir of water in form of remnant glacier deposits concentrated on mountain flanks. Geologic mapping (Head and Marchant, 2003; Head et al., 2005; Levrard et al., 2004; Levy et al., 2014; Shean et al., 2005) suggests the area of these deposits is at least $7.1 \times 10^5$ km$^2$ but the overall volume and exchange rates of this reservoir are only poorly constrained. Therefore, in this model, we assume an initial volume for this reservoir that provides a sufficient source for the present-day NPLD.

In a pioneering attempt to calculate the D/H variations of the polar cap, Fisher (2007) considered an atmospheric reservoir and tracked its isotopic evolution. This is appropriate if the atmospheric mixing time is comparable to the characteristic exchange times in the model. However, because the expected amount of water exchanged between the reservoirs on kyr timescales (Levrard et al., 2007) is several orders of magnitude greater than the atmospheric water content (Smith, 2008), the atmosphere can be regarded as thin conduit through which the reservoirs exchange. Thus, the atmospheric reservoir is negligible for the timescale under consideration, and so is the fractionation at the top of the atmosphere to space. For the long-term (Gyr) evolution atmospheric loss must be considered (Kass and Yung, 1999); for seasonal effects the atmospheric circulation dynamics are essential (Fisher, 2007; Krasnopolsky, 2015; Novak et al., 2002).

This central aspect of our model warrants additional discussion. Cloud micro-physics can play an important role in the Mars water cycle (Montmessin et al., 2004; Richardson et al., 2002; Wang and Ingersoll, 2002), allowing fractionation upon condensation in the atmosphere relative to direct deposition on the surface. This effect may be significant on short timescales, but it can be neglected on long timescales for the following reason. As the atmosphere is negligible in total water content relative to the integrated fluxes over kyr timescales, even if temporary fractionation occurs at the clouds which subsequently deposit, this diurnal/annual effect cannot build up over many cycles. The total atmospheric input must equal the total output on long timescales as the water molecules pass through the atmosphere. To the (limited) extent that clouds migrate across latitudes, they can transport with them an isotopic anomaly, but only as long as the residual water in the atmosphere also condenses to remove the complement of that anomaly to another reservoir every cycle. This mechanism should be investigated quantitatively by more detailed GCMs in the future, but it too, is limited, because cloud migration is confined in latitude by the boundaries of the overturning convection cells, seen in models and observations (Montmessin et al., 2004; Richardson et al., 2002; Wang and Ingersoll, 2002).

Interannual variations of the atmospheric humidity occur on Mars, such as during the 1969 southern summer when a factor of three increase in atmospheric water content was observed (Jakosky and Barker, 1984). However, such variations would only be important here if they repeated for a substantial fraction of the 1 kyr integration time, and thus change the average humidity (as may occur if the CO2 cover of Southern cap is lost for extended periods).

The implementation of the model assumptions as outlined is described below.

**Model Description**

We construct a model which tracks the transfer of $H_2O$ and HDO among the relevant reservoirs on Mars, and calculates the resulting abundances. Standard values are used for the obliquity, eccentricity and longitude of perihelion as function of time (Laskar et al., 2004). Each of the reservoir fluxes is computed independently as follows.

*Polar Regions:* The polar flux is taken from previous GCM-based model results (Levrard et al., 2007), where ice accumulation at the surface includes a constant dust fraction. Ice sublimation is also assumed to occur at the top-most layer. Some heterogeneous structure does exist (Smith et al., 2018), such as possible enhancement in the loss rate at troughs. We neglect these variations here because trough retreat is expected to be limited by growth of a sublimation lag that dramatically reduces the flux (Bramson et al., 2018). This assumption is supported by the presence of lag layers that are extensive and consistent across the cap (Milkovich and Head, 2005).

We implement a modification in the ablation rate that accounts for the dependence of the loss rate on the thickness of the growing lag (Hudson et al., 2007). The diffusion barrier of the sublimation lag alters the sublimation flux $J_{\text{NPLD}}$ in our model, from its full value $J_0$ according to

$$J_{\text{NPLD}} = \frac{1}{1+z/z_0} J_0, \qquad (1)$$

where $z$ is the dust layer thickness above the ice, $z_0$ is a variable parametrizing the rate of reduction of flux with depth, and $J_0$ is the flux taken from (Levrard et al., 2007). This expression reduces the flux from full to half its value from $z = 0$ to $z_0$ respectively, and approaches 0 as $z$ grows large.

This improvement relative to past models (Levrard et al., 2007) affects the thickness of dust lag layers. With a constant diffusion inhibition factor, the dust layer thickness distribution would have a tail with large thicknesses.

Layer formation in the cap is tracked, as is the D/H ratio of each layer. Ice loss or gain is assumed to occur at the top-most layer. Equilibrium fractionation (Merlivat and Nief, 1967) is assumed at deposition, while ablation is non-fractionating. We ensure layers of all thicknesses are resolved by following their boundaries explicitly during the simulation. There is no spatial discretization of the layers; instead their boundaries are tracked at each time step.

*Subsurface ice deposits:* Atmospheric humidity is prescribed here by simulations of the global circulation as function of obliquity (Schorghofer and Forget, 2012). The SSD flux is computed from the migration rate of the latitudinal margin of the ice table. Previous modeling work by Schorghofer and Forget (2012) has shown that the change in ground ice volume is dominated by changes at a latitude range where the ice table depth increases rapidly, and not by the small changes in ice table depth that occur at more poleward latitudes. For this reason, the volume change is formulated in terms of the movement of an imaginary point in latitude. This stability margin is calculated from a 1D thermal model (Schorghofer, 2008), by equating the mean saturation water vapor density in the subsurface with the mean atmospheric vapor density. For simplicity, we assume the thickness of the SSD is constant. This thickness does not necessarily correspond to the true thickness of the SSD, but represents the depth of the exchangeable reservoir. The SSD is stratified horizontally, and the D/H ratio in latitude bands is tracked, with analogous assumptions to the cap.

Fractionation due to adsorption in the regolith (Moores et al., 2011) can also alter the dynamics of exchange with the SSD. However, this adsorptive reservoir is orders of magnitude smaller than the typical SSD reservoir (of one meter thickness extending to 40º latitude) (Jakosky, 1983). Isotopic differences in the diffusion coefficient were measured (Moores et al., 2011), and while the measurement accuracy allows only approximate differences to be reported, these differences cannot affect the overall isotopic budget when the amount lost is significantly greater than the adsorptive capacity.

*Global system:* Lastly, the equatorial glacier flux $J_{GD}$ is obtained from mass conservation,

$$J_{NPLD} + J_{SSD} + J_{GD} = 0. \qquad (2)$$

The temperature controlling the kinetic fractionation factor is assumed to be the mean annual temperature, and calculated from the thermal model as a function of latitude and orbital parameters (Schorghofer, 2008). For ice accumulation, the fractionation factor (Merlivat and Nief, 1967) α is given by

$$\alpha = \exp\left(\frac{16288}{T^2} - 0.0934\right), \qquad (3)$$

where T is the temperature at deposition.

The latitude taken for the temperature to evaluate α is 85° for the NPLD, the instantaneous margin's latitude for the SSD, and 20° for the GD. While somewhat arbitrary, we verified the precise choice of representative latitude does not significantly alter the results. The HDO deposition flux of the gaining reservoirs is then apportioned according the relative fluxes and $\alpha(T)$, according to

$$J_i^{HDO} = \frac{\alpha(T_i)J_i}{\sum_j^+ \alpha(T_j)J_j}\left(\sum_j^- J_j\right). \qquad (4)$$

where $\Sigma^+$ and $\Sigma^-$ refer to sums over the gaining and receding reservoirs, respectively, and the index can be NPLD, SSD, or GD. Two simultaneous sinks are required to cause D/H fractionation among the ice reservoirs. When only one reservoir grows, no fractionation occurs, and the HDO flux of the accumulated ice is simply the sum of the two losing sources. Hence, in a two-reservoir model, fractionation is not possible because molecules simply transfer back and forth between the source and sink.

The model is initialized with no polar cap, and a 17 m GEL of GD at 21 Myr ago. This initial value does not affect the model dynamics as long as it is sufficient to supply the net growth of the cap. The initial SSD latitudinal margin corresponds to the initial humidity, and for a 1m thick reservoir results in 0.37 m GEL of ice. The initial D/H ratio of the reservoirs is arbitrary in our model.

**Results**

Figure 2 shows NPLD, SSD, and GD fluxes and the obliquity (axial tilt) as function of time for the last 8 Myr. Our nominal model parameters were chosen to be a constant 2% dust content as suggested by MARSIS (Picardi et al., 2005), and $z_0$=5 cm (see Methods), resulting in ~2200 m current cap thickness. The thickness of the SSD was chosen to be 1 m of exchangeable ice (the sensitivity to these parameters is examined later). Obliquity exerts a strong influence over the individual fluxes. Following previous predictions (Levrard et al., 2007), NPLD growth occurs when the obliquity drops below a threshold level, which last occurred ~3.2 Myr ago.

Figure 3a plots the time evolution of the reservoir sizes, showing the oscillations and net growth of the NPLD. While the SSD margin oscillates, net polar growth occurs at the expense of the GD on long timescales. However, on layer formation timescales, all three reservoirs participate in the exchange. During times of small and relatively constant obliquity, such as between 2.7-2.2 Myr and from 0.4 Myr to present, the cap grows significantly. At these low obliquity values, the cap experiences no intermittent periods of sublimation.

Figure 3b compares the evolution of the SSD margin to that of the obliquity. During periods when the cap grows monotonically, there are still variations in the location of the margin of the SSD, driven by 50 kyr precession cycles. This variation in flux from the mid-latitudes is captured in the isotopic record of the polar deposits, despite the polar accumulation rate remaining approximately constant.

A simulated profile of the polar deposits as a function of depth is shown in Figure 4, as an ice core might appear. Alternations of ice and dust-rich layers are visible, as the physical manifestations of accumulation/sublimation rate variations. We see again two regimes: during times of low and almost constant obliquity the cap grows significantly, and times with sublimation periods that give rise to unconformities and lag layers. The thickness of the lag layers depends on the varying sublimation rate, while their separation corresponds to the roughly constant obliquity period of ~120 kyr. In addition to this previously studied (Hvidberg et al., 2012) physical stratigraphy, the isotopic record reveals further aspects of the dynamics of the water cycle such as the source reservoir and temperature differences between reservoirs. The D/H profile is also shown in Figure 4, and it exhibits oscillations of an amplitude of ~10% about the nominal SMMW. The D/H oscillations are not symmetric. During times of NPLD ice accumulation following cap recession, the D/H ratio shows an initial value lower than the remains of the layer below. Thus, sublimation layers in the cap are predicted to have a characteristic increase in D/H across the layer from top to bottom.

We use Lomb-Scargle Periodogram for unevenly sampled data to analyze the periodicity of the accumulation rate. The periodogram shows a peak at a period of 120 kyr and higher harmonics starting at 60 kyr (Figure 5a). These harmonics appear due to the unconformities in the record during sublimation intervals. To better understand the spectrum, we window the data (Figure 5b) with a running window of 600 kyr. Regions without dominant periodicity are seen at times of low and almost constant obliquity where the cap grows monotonically, and regions where the dominant period is ~120 kyr are seen when the obliquity exceeds a critical value (Levrard et al., 2007).

Using the same spectral technique to analyze the D/H profile, an additional peak emerges at 50 kyr (Figure 5c). We again window the data in depth (Figure 5d), and find the 120 kyr obliquity cycle is still evident where the dust layers are present. However, at epochs of low and almost constant obliquity, such as at present, the 50 kyr precession cycles appear in the isotope signal. The oscillation is a source effect—it is inherited from the oscillations in the SSD margin, which is sensitive to the precession cycle, even when the growing NPLD is not (Schorghofer, 2008). This result demonstrates that different climate processes may be dominant at different epochs.

We tested the sensitivity of the model to assumed parameters. Lag diffusion parameters and dust fraction affect the model similarly (Figure 6a). Both alter the net cap growth rate, but do not strongly influence its isotopic composition. Greater dust fraction and smaller value of $z_0$ both inhibit diffusive loss and thus enhance cap growth. At low dust content, the model results converge to one value, as expected. At small values of $z_0$, even a thin sublimation lag reduces the flux substantially, and hence the cap grows to a height greater than that observed today. The variations in cap growth rate are also reflected in the predicted ice layer thicknesses (not in additional layers).

The assumed SSD thickness affects the D/H variation of the layers at thicknesses of a few meters or smaller (Figure 6b). The dependency is weaker when the SSD is thicker. This result may be understood as the typical polar water flux predicted by GCMs (~10's pr μm/year) is equivalent in magnitude to latitudinal oscillations in the SSD when its thickness is ~1.5 m. At smaller thicknesses the SSD directly exchanges with the polar cap and hence affects the polar D/H record.

At larger thicknesses, the greater flux is accommodated in the model by the third, non-polar reservoir, and the polar isotopic record shows little sensitivity to the enhanced flux.

**Conclusions**

We have constructed a simplified three box model which tracks the transfer of $H_2O$ and HDO between three major ice reservoirs and calculates the resulting abundances, and their dependence on assumed parameters, over the period of the NPLD-buildup. The diffusion parameters and dust content in the cap influence the overall ice cap size and layer thickness. We find that a 2% ice/dust ratio, as suggested by MARSIS (Picardi et al., 2005), is compatible with growth of a ~2-km thick cap. Unlike past models, we predict the D/H variation of the NPLD is dominated by the temperature difference between depositional reservoirs and the SSD thickness. For our nominal parameters, the amplitude of the predicted D/H variation is comparable to that in terrestrial ice cores and measurable by future missions. Most of the time, the latitudinal extent of the SSD is controlled by obliquity, but periods of relatively constant obliquity record 50 kyr precession cycles. At these times of relatively low and constant obliquity, as at present, the NPLD continuously grows from year to year, while the SSD margin location oscillates at the precession period. Hence, isotopic sampling of the top 100 m may reveal climate oscillations unseen in the layer thicknesses and should probe the recent precession cycle. Future models can improve our predictions by taking into account more accurate climate model variations at past orbital configurations, especially in regards to the annual humidity cycle at high and low obliquity, as well as the role of cloud condensation and migration in fractionating the hydrogen reservoirs.

**Acknowledgements**

The authors wish to thank the Minerva Center for Life Under Extreme Planetary Conditions, the I-CORE Program of the PBC and ISF (Center No. 1829/12) and the Helen Kimmel Center for Planetary Sciences for support of this work. We also thank Dr. Ivy Curren for her assistance with the graphics.

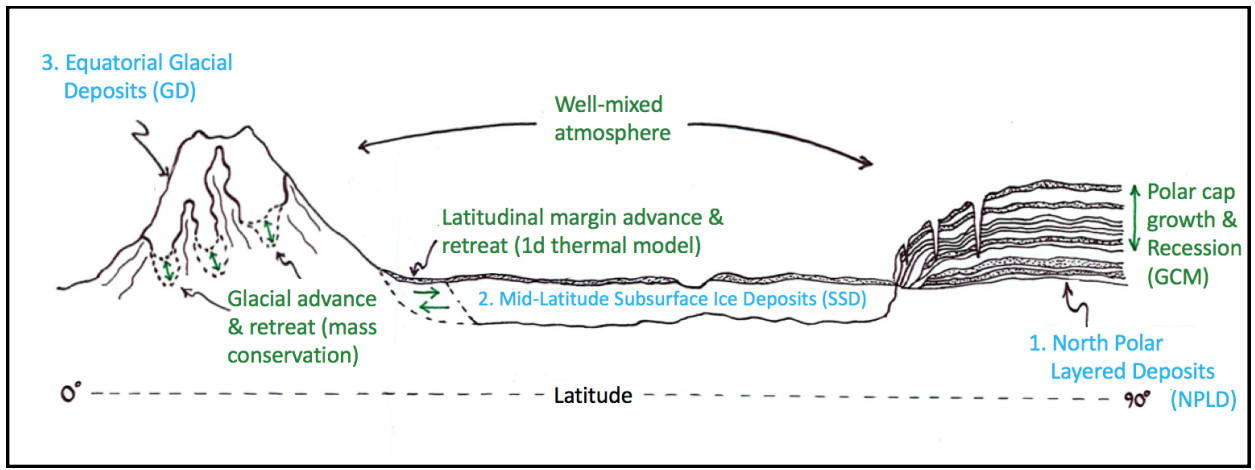

**Fig.1: Schematic illustration of the three-box model.** The primary reservoirs are shown at the polar (NPLD), mid- (SSD), and low- (GD) latitudes.

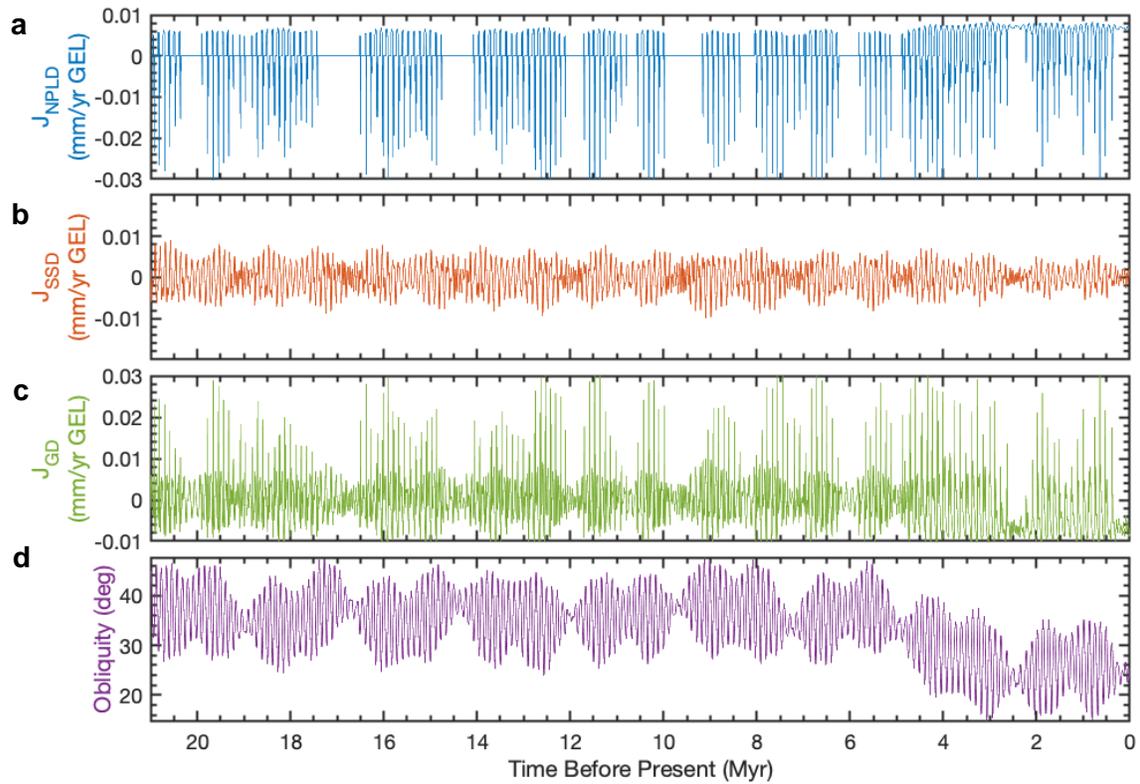

**Fig. 2: Evolution of the reservoir fluxes and obliquity.** Fluxes are in units of millimeter Global Equivalent Layer (GEL) thickness per year, positive (negative) for a growing (receding) reservoir. **a,** North Polar Layered Deposits, **b,** Subsurface Deposits, **c,** Glacial Deposits, **d,** Obliquity.

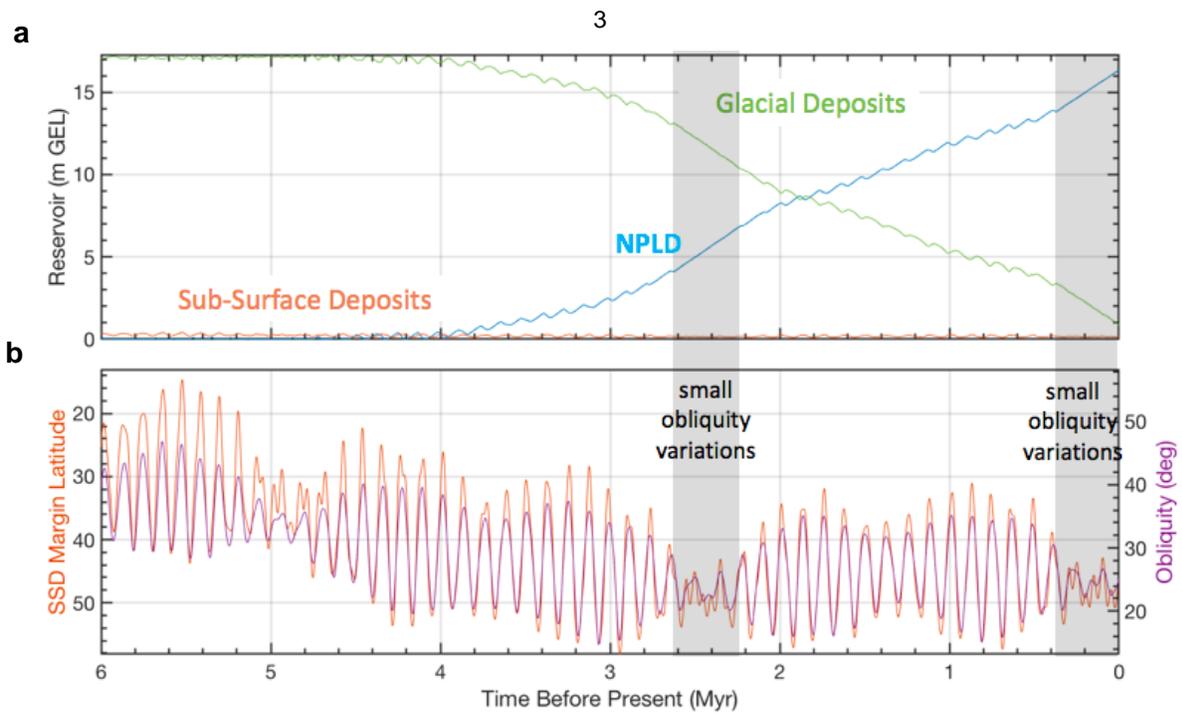

**Fig. 3: Evolution of integrated reservoir sizes. a,** While the SSD margin oscillates, net polar growth ultimately occurs at the expense of the GD. On layer formation timescales, all three reservoirs participate in the exchange. **b**, SSD margin latitude (red) and obliquity (purple) as a function of time. The margin follows the obliquity except at times when the obliquity value is low and almost constant (shaded in grey).

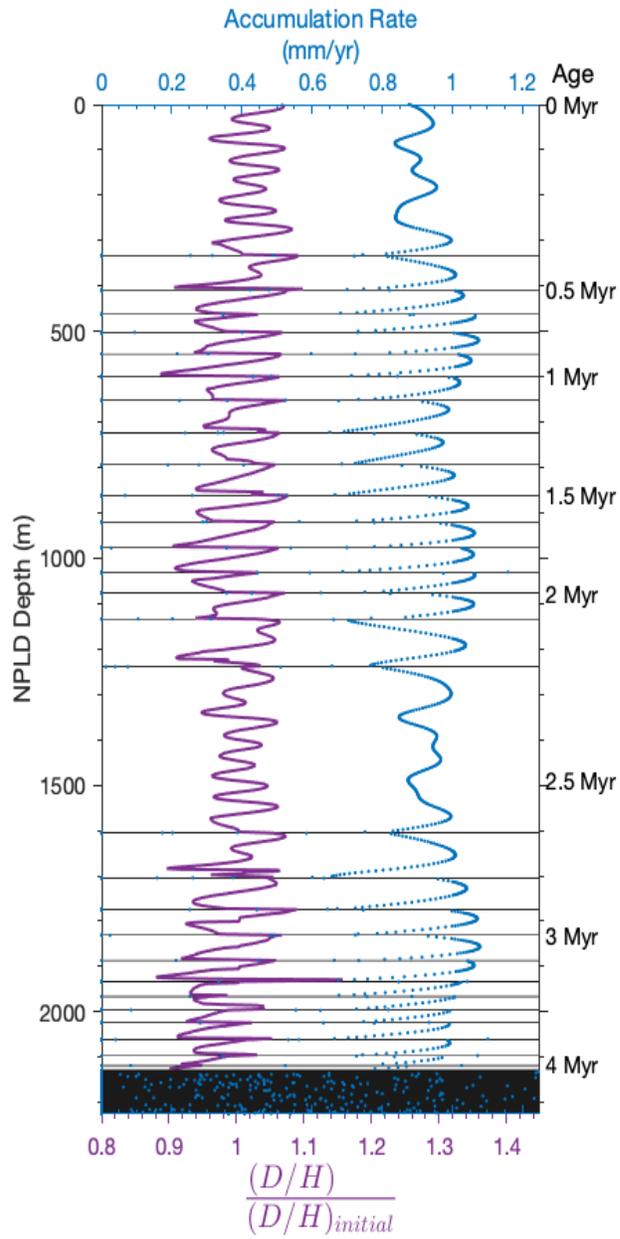

**Fig. 4: Profile of the polar deposits as function of depth and time.** White layers represent deposited ice and grey are dust lags. The accumulation rate (blue) and the D/H ratio (purple) as function of depth are shown for the nominal model parameters.

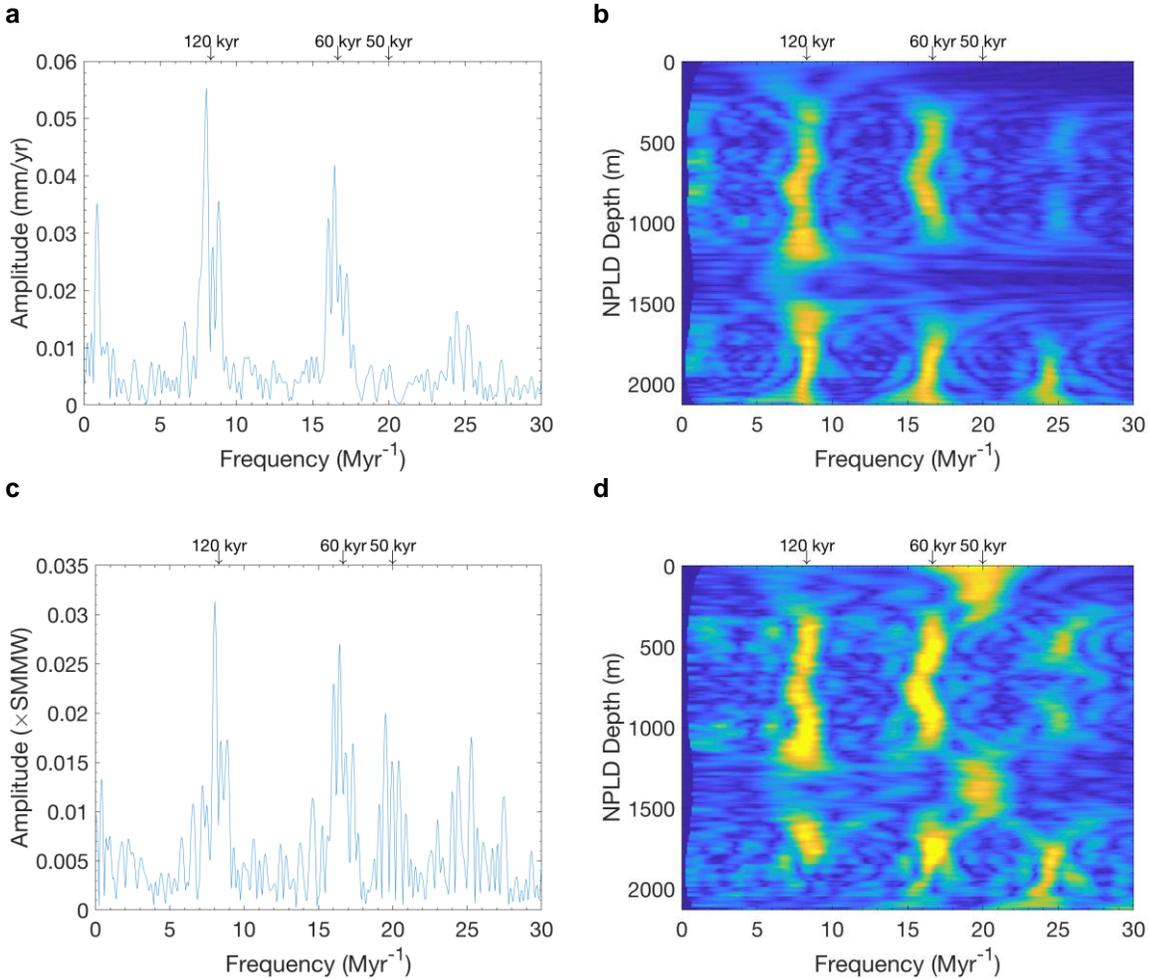

**Fig. 5: Lomb-Scargle periodograms highlighting dominant frequencies in the polar record. a,** Accumulation rate periodogram shows a peak at 120 kyr and higher harmonics, as well as the ~1.1 Myr period envelope of the obliquity variations. **b,** Windowed periodogram (window size 300 m) with depth shows regions with no dominant periodicity at times of low and almost constant obliquity (such as 2.7 to 2.2 Myr and from 0.4 Myr to present) and regions where the dominant period is ~120 kyr. **c,** A periodogram of the D/H signal shows that in addition to the obliquity peaks, a new peak appears at 50 kyr. **d,** Windowed periodogram (window size 300 m) with depth for the D/H signal shows that at epochs of low and almost constant obliquity (as above), the 50 kyr precession cycles is dominant.

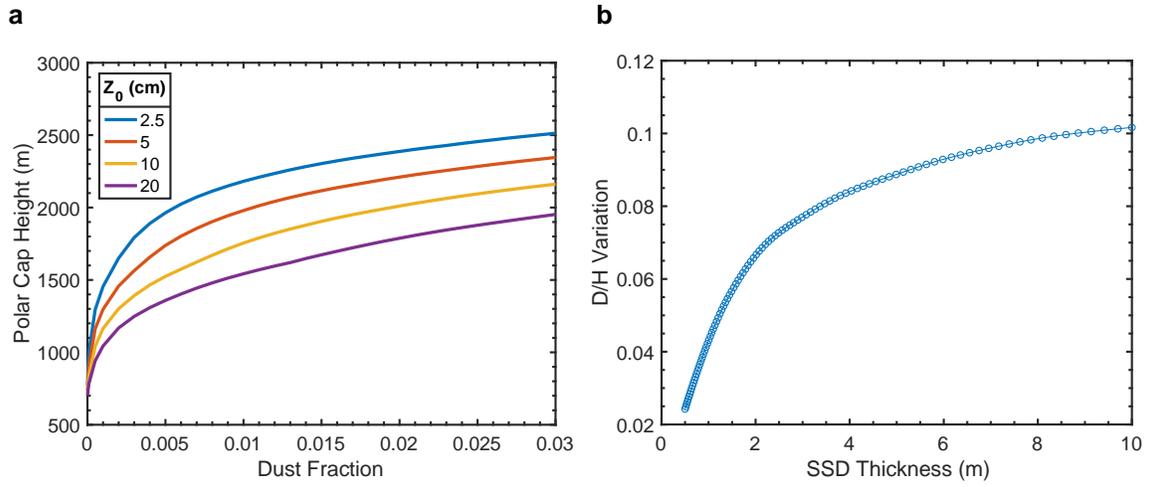

**Fig. 6: Sensitivity to assumed model parameters. a,** Polar cap height as function of dust fraction and diffusion parameters **b,** Amplitude of the D/H variation increases with assumed SSD thickness